\documentclass[fleqn,twoside]{article}
\usepackage{espcrc2}

\usepackage{graphicx}
\usepackage[figuresright]{rotating}

\newcommand{\AmS}{{\protect\the\textfont2
  A\kern-.1667em\lower.5ex\hbox{M}\kern-.125emS}}
\newcommand{\beq}{\begin{equation}}
\newcommand{\eeq}{\end{equation}}
\newcommand{\bea}{\begin{eqnarray}}
\newcommand{\eea}{\end{eqnarray}}

\def\proof{\noindent{\sl Proof:}\kern0.6em}

\def\frac#1#2{\hbox{$#1\over#2$}}
\def\dual{\mathstrut^*\kern-0.1em}

\def\lvec#1{\setbox0=\hbox{$#1$}
    \setbox1=\hbox{$\scriptstyle\leftarrow$}
    #1\kern-\wd0\smash{
    \raise\ht0\hbox{$\raise1pt\hbox{$\scriptstyle\leftarrow$}$}}
    \kern-\wd1\kern\wd0}
\def\rvec#1{\setbox0=\hbox{$#1$}
    \setbox1=\hbox{$\scriptstyle\rightarrow$}
    #1\kern-\wd0\smash{
    \raise\ht0\hbox{$\raise1pt\hbox{$\scriptstyle\rightarrow$}$}}
    \kern-\wd1\kern\wd0}

\def\nabstar#1{\nabla\kern-0.5pt\smash{\raise 4.5pt\hbox{$\ast$}}
               \kern-4.5pt_{#1}}

\def\drvstar#1{\partial\kern-0.5pt\smash{\raise 4.5pt\hbox{$\ast$}}
               \kern-5.0pt_{#1}}

\def\momp#1#2{
    \setbox0=\hbox{${#1}$}\setbox1=\hbox{${#1}_{#2}$}
    {#1}_{#2}\kern-\wd1\kern\wd0
    \smash{\raise4.5pt\hbox{$\scriptscriptstyle +$}}}
\def\momm#1#2{
    \setbox0=\hbox{${#1}$}\setbox1=\hbox{${#1}_{#2}$}
    {#1}_{#2}\kern-\wd1\kern\wd0
    \smash{\raise4.5pt\hbox{$\scriptscriptstyle -$}}}
\def\mompm#1#2{
    \setbox0=\hbox{${#1}$}\setbox1=\hbox{${#1}_{#2}$}
    {#1}_{#2}\kern-\wd1\kern\wd0
    \smash{\raise4.5pt\hbox{$\scriptscriptstyle \pm$}}}
\def\smomp#1#2{
    \setbox0=\hbox{${#1}$}\setbox1=\hbox{${#1}_{#2}$}
    {#1}_{#2}\kern-\wd1\kern\wd0
    \smash{\raise3pt\hbox{$\scriptscriptstyle +$}}}
\def\smomm#1#2{
    \setbox0=\hbox{${#1}$}\setbox1=\hbox{${#1}_{#2}$}
    {#1}_{#2}\kern-\wd1\kern\wd0
    \smash{\raise3pt\hbox{$\scriptscriptstyle -$}}}
\def\smompm#1#2{
    \setbox0=\hbox{${#1}$}\setbox1=\hbox{${#1}_{#2}$}
    {#1}_{#2}\kern-\wd1\kern\wd0
    \smash{\raise3pt\hbox{$\scriptscriptstyle \pm$}}}
\def\si{\kern1pt{\rm si}}
\def\co{\kern1pt{\rm co}}

\def\psibar{\bar{\psi}}

\def\psiprime{\psi\kern1pt'}
\def\psibarprime{\psibar\kern1pt'}
\def\rhoprime{\rho\kern1pt'}
\def\rhobar{\bar{\rho}}
\def\rhobarprime{\rhobar\kern1pt'}
\def\rhobartilde{\kern2pt\tilde{\kern-2pt\rhobar}}
\def\rhobartildeprime{\kern2pt\tilde{\kern-2pt\rhobar}\kern1pt'}

\def\zetabar{\bar{\zeta}}
\def\zetaprime{\zeta\kern1pt'}
\def\zetabarprime{\zetabar\kern1pt'}
\def\zetar{\zeta_{\raise-1pt\hbox{\sixrm R}}}
\def\zetabarr{\zetabar_{\raise-1pt\hbox{\sixrm R}}}

\def\phiimpr{\phi_{\kern0.5pt\hbox{\sixrm I}}}

\def\diracstar#1#2{
    \setbox0=\hbox{$\gamma$}\setbox1=\hbox{$\gamma_{#1}$}
    \gamma_{#1}\kern-\wd1\kern\wd0
    \smash{\raise4.5pt\hbox{$\scriptstyle#2$}}}

\def\f1{f_1}

\def\h1{h_1}

\def\opprime#1{\setbox0=\hbox{${\cal O}$}\setbox1=\hbox{${\cal O}_{\rm #1}$}
    {\cal O}_{\rm #1}\kern-\wd1\kern\wd0
    \smash{\raise4.5pt\hbox{\kern1pt$\scriptstyle\prime$}}\kern1pt}

\def\ophatprime#1{\setbox0=\hbox{$\widehat{\cal O}$}
    \setbox1=\hbox{$\widehat{\cal O}_{\rm #1}$}
    \widehat{\cal O}_{\rm #1}\kern-\wd1\kern\wd0
    \smash{\raise4.5pt\hbox{\kern1pt$\scriptstyle\prime$}}\kern1pt}

\def\bopprime#1{\setbox0=\hbox{${\cal O}$}\setbox1=\hbox{${\cal O}_{\rm #1}$}
    {\cal L}_{\rm #1}\kern-\wd1\kern\wd0
    \smash{\raise4.5pt\hbox{\kern1pt$\scriptstyle\prime$}}\kern1pt}

\def\blagprime#1{\setbox0=\hbox{${\cal B}$}\setbox1=\hbox{${\cal B}_{#1}$}
    {\cal B}_{#1}\kern-\wd1\kern\wd0
    \smash{\raise5.2pt\hbox{\kern1pt$\scriptstyle\prime$}}\kern1pt}

\def\msbar{{\rm \overline{MS\kern-0.05em}\kern0.05em}}

\newcommand{\bes}{\begin{eqnarray}}
\newcommand{\ees}{\end{eqnarray}}

\title{Twisted mass Lattice QCD\thanks{Based on a plenary talk given at
Lattice 2004, the
XX International Symposium on Lattice Field Theory, held on June, 21--26, 2004
at Fermilab, Batavia, IL, U.S.A.} 
      }

\author{R. Frezzotti\address[infnmi]{I.N.F.N. Milano and University 
        of Milano Bicocca, 
        Piazza della Scienza 3, I-20126 Milano, Italy}
        }
       
\begin{document}

\begin{abstract}
I review the main theoretical properties and some recent
analytical and numerical investigations of the formulations
of lattice QCD with chirally twisted Wilson quarks (also
known as twisted mass lattice QCD). 
%
%
\vspace{1pc}
\vspace{-0.3cm}
\end{abstract}

\maketitle
\begin{flushright}
Preprint Bicocca-FT-04-13
\end{flushright}

In this contribution I review lattice formulations based on
so-called twisted mass (tm) Wilson quarks, focusing 
on the general properties and problems that in my opinion are most relevant for 
on-going studies and applications to physics~\footnote{The reader is urged to consult
the cited papers for more details. For several aspects this write-up 
should be regarded as an introduction (or a comment) to those works.}.

The simplest of these 
formulations, which corresponds to QCD with one mass degenerate quark doublet, 
was introduced in ref.~\cite{TM} as a way to get rid of the spurious quark zero 
modes that plague standard Wilson fermions and is referred to as 
twisted mass lattice QCD (tmLQCD).

\section{$N_f=2$ twisted mass lattice QCD}
\label{SECT2}

Following refs.~\cite{TM,F_02,FR1}, a sensible lattice formulation of QCD 
with an $SU_f(2)$ flavour doublet of mass degenerate quarks is 
given by the action $S = S_{g}[U] + S_F^{(\omega)}[\psi,\bar\psi,U]$,
where $S_{g}[U]$ stands for any discretization of the YM action and
\begin{eqnarray}
&&S_{F}^{(\omega)}[\psi,\bar\psi,U]=
a^4 \sum_x\,\bar\psi(x)
\Big{[}\gamma\cdot\widetilde\nabla+
\nonumber \\
&&+e^{-i\omega\gamma_5\tau_3}
W_{\rm{cr}}(r)+m_q\Big{]}\psi(x)\, , \label{PHYSCHI}
\end{eqnarray}
\vspace{-0.4cm}
\begin{eqnarray}
&&\gamma\cdot\widetilde\nabla\equiv
\frac{1}{2}\sum_\mu\gamma_\mu(\nabla^\star_\mu+\nabla_\mu)\, ,\label{WDONDD}
\\
&&W_{\rm{cr}}(r)\equiv-a\frac{r}{2}\sum_\mu\nabla^\star_\mu\nabla_\mu+
M_{\rm{cr}}(r) \, . \label{WDONDM}
\end{eqnarray}
Here $M_{\rm{cr}}$ denotes the critical mass, 
$m_q$ is the (yet unrenormalized) offset quark mass (such that,
once $M_{\rm{cr}}$ has been set to the appropriate value, $m_\pi^2 
\sim m_q$ as $m_q \to 0$), while $r\in[-1,1]$ and $\omega\in[-\pi,\pi)$ 
characterize
the specific Wilson-type UV regularization.
As discussed below (see sect.~\ref{sec:MC}), $M_{\rm{cr}}$ is,
up to O($a$) uncertainties, independent of $\omega$. With the whole O($a$)
and numerical uncertainty on the critical point brought onto $M_{\rm{cr}}(r)$,
one can treat $\omega$ and $m_q$ (besides $r$) as exactly known parameters. 

{\small \vspace{0.2cm}

For $\omega=0$ the familiar action for LQCD with two mass 
degenerate Wilson quarks is recovered.
As long as $\omega \neq 0$ and $m_q \neq 0$, no zero modes of the
(two-flavour) Dirac 
operator in eq.~(\ref{PHYSCHI}), which I denote by $D_F$, can occur
on any gauge configuration, because~\cite{AOGO}
\begin{eqnarray}
&& {\rm Det} [ D_F ] = {\rm det} [Q^2 + m_q^2 \sin^2\omega ] \, ,
\label{DET2F} \\
&& Q \equiv \gamma_5 [\gamma \!\cdot\! \widetilde{\nabla}
          + W_{\rm cr}(r) + m_q \cos \omega ]  = Q^\dagger
\label{QDEF} \, .
\end{eqnarray}
The property~(\ref{DET2F}) solves the problems related to
exceptional configurations in the quenched approximation
as well as to MD instabilities due
to exceptionally small eigenvalues of $Q^2$ in HMC-like
algorithms for unquenched simulations.

The extension of unquenched simulation algorithms of the HMC and  multiboson type
to the action~(\ref{PHYSCHI}) --or eq.~(\ref{TWISCHI}) below-- 
is straightforward~\cite{FARCH}. Preliminary
results from on-going simulations of tmLQCD seem to hint at a numerical
cost comparable to that of simulations with staggered fermions~\cite{KE_PROC}.

} 

\subsection{Symmetries and reflection positivity}
\label{SRP}

The symmetries of $N_f=2$ tmLQCD are discussed in detail 
in refs.~\cite{TM,FR1}. Here I recall only the main points
arising from those analyses.

{\small \vspace{0.2cm}

The chirally twisted Wilson term in the action~(\ref{PHYSCHI}) breaks 
flavour chiral symmetry in such a way that, if $m_q=0$, three symmetry 
generators are preserved (see~\cite{FR1}, eq.~(4.7)) for any $\omega$-value.
Correspondingly, there exist three exactly conserved isotriplet 
lattice currents (see~\cite{FR1}, eqs.~(4.8)--(4.9)). One of
them is the neutral vector current, while the other two are mixtures of
the charged axial and vector currents (they get purely axial at $\omega =
\pm \pi/2$ and vector at $\omega=0,\pi$). 

Charge conjugation symmetry remains exact, but, owing
to the chiral twisting of the Wilson term,
all single-axis inversions leave the action invariant
only if combined with the discrete isospin rotations ${\cal T}_{1,2}$
\beq
{\cal T}_{1,2}: \;
\psi(x) \to i\tau_{1,2} \psi(x) \, , \;\;
\bar\psi(x) \to -i \bar\psi(x) \tau_{1,2} \, .
\label{DIISRO}
\eeq
or, alternatively, a sign change of the twist angle $\omega$.
In particular this remark holds for the physical parity 
transformation (here $x_P \equiv (-{\bf {x}},t)$)
\begin{eqnarray}
{\mathcal {P}}:\left \{\begin{array}{ll}
&\hspace{-.3cm}U_0(x)\rightarrow U_0(x_P) \\
&\hspace{-.3cm}U_k(x)\rightarrow U_k^{\dagger}(x_P-a\hat{k})\, , \quad k=1,2,3\\
&\hspace{-.3cm}\psi(x)\rightarrow \gamma_0 \psi(x_P)\\
&\hspace{-.3cm}\bar{\psi}(x)\rightarrow\bar{\psi}(x_P)\gamma_0 \quad .
\end{array}\right . \label{PAROP}
\end{eqnarray}
For later use I note the non-anomalous symmetry~\cite{FR1} 
\beq
{\cal R}_5^{d} \equiv {\cal R}_5 \times {\cal D}_d \, ,
\label{R5Dd}
\eeq
where
\beq
{\mathcal R}_5 :\left \{\begin{array}{ll}
\hspace{-.3cm}&\psi(x)\rightarrow \gamma_5 \psi(x)  \\
\hspace{-.3cm}&\bar{\psi}(x)\rightarrow - \bar{\psi}(x) \gamma_5 \, ,
\end{array}\right  . \label{R5}
\eeq
\beq
{\mathcal{D}}_d : \left \{\begin{array}{lll}
\hspace{-.3cm}&U_\mu(x)\rightarrow U_\mu^\dagger(-x-a\hat\mu) \\
\hspace{-.3cm}&\psi(x)\rightarrow e^{3i\pi/2} \psi(-x)  \\
\hspace{-.3cm}&\bar{\psi}(x)\rightarrow e^{3i\pi/2} \bar{\psi}(-x) \, .
\end{array}\right . \label{Dd}
\eeq

\vspace{0.0cm} } 

Correlation functions evaluated with the tmLQCD action enjoy link reflection
positivity for all values of $r\in[-1,1]$ and $\omega \in [-\pi,\pi)$~\cite{IY}, 
as well as site reflection positivity provided $|r|=1$ and
$|8r + 2aM_{\rm cr}(r) + 2am_q \cos\omega| >6 $~\cite{TM,FR1}.


\subsection{Renormalizability}
\label{sec:RE}

The tmLQCD fermionic action~(\ref{PHYSCHI}) is written in what is usually
called the ``physical basis''~\cite{FR1}, with $m_q$ real
(and positive). By a change of basis,
\beq
\chi = e^{-i\omega\gamma_5\tau_3/2} \psi \, , \quad\quad
\bar\chi = \bar\psi e^{-i\omega\gamma_5\tau_3/2}  \, , \label{TWCHI}
\eeq
the action takes the form considered in
ref.~\cite{TM},
\begin{eqnarray}
&& S_{F}^{(\omega)}[\chi,\bar\chi,U] \; = \;
a^4 \sum_x\,\bar\chi(x)
\Big{[}\gamma\cdot\widetilde\nabla + 
\nonumber \\
&&
-a\frac{r}{2}\sum_\mu\nabla^\star_\mu\nabla_\mu+ m_0 
+ i m_2\gamma_5\tau_3 \Big{]}\chi(x)\, ,
\label{TWISCHI}
\end{eqnarray}
\beq
\quad\; m_0 = M_{\rm{cr}} + m_q \cos\omega \, , \quad\;
m_2 = m_q \sin\omega \, . \label{BQMP}
\eeq
{\small \vspace{0.1cm}
All the symmetry properties described above or in 
refs.~\cite{TM,FR1} can be expressed in the
quark basis~(\ref{TWCHI}). In particular I note the 
spurionic symmetries of the action~(\ref{TWISCHI})
\beq
{\cal R}_5 \times (r \to -r) \times (m_0 \to -m_0)
\times (m_2 \to -m_2) \, ,   \label{R5_sp2}
\eeq
\vspace{-0.1cm}
\beq
\tilde{P} \times (m_2 \to -m_2) \, ,
\label{TILPAR_sp} 
\eeq
with ${\cal R}_5$ form invariant under~(\ref{TWCHI}) and
\begin{eqnarray}
\tilde{\mathcal P}:\left \{\begin{array}{ll}
&\hspace{-.3cm}U_0(x)\rightarrow U_0(x_P) \\
&\hspace{-.3cm}U_k(x)\rightarrow U_k^{\dagger}(x_P-a\hat{k})\, , \quad k=1,2,3\\
&\hspace{-.3cm}\chi(x)\rightarrow \gamma_0 \chi(x_P)\\
&\hspace{-.3cm}\bar{\chi}(x)\rightarrow\bar{\chi}(x_P)\gamma_0 \quad .
\end{array}\right . \label{TILPAROP}
\end{eqnarray}

Standard symmetry and power counting arguments imply~\cite{TM} that
(up to terms irrelevant as $a \to 0$) the most general action for 
two-flavour tmLQCD is just of the form $S_g[U] + S_{F}^{(\omega)}
[\chi,\bar\chi,U]$, with $S_g[U] \propto 1/g_0^2$ and $S_{F}^{(\omega)}$ 
given in eq.~(\ref{TWISCHI}). 
Invariance under ${\cal P} \times {\cal T}_{1,2}$ (see eqs.~(\ref{PAROP})
and~(\ref{DIISRO})) rules out parity odd pure gauge terms
($\propto tr[ F \widetilde{F} ]$ as $a \to 0$). 
While $g_0^2$ and $m_2$ need only multiplicative renormalization,
$m_0$ undergoes additive and multiplicative renormalization~\cite{TM}.

\vspace{0.2cm} } 

The continuum flavour chiral WTI's of QCD with two mass degenerate quarks 
can be implemented (up to O($a$)) for all 
$\omega$-values~\cite{TM,FR1}, with the renormalized current quark mass given by 
\begin{eqnarray}
& \hat{m}_q = Z_P^{-1}Z_M(\omega) m_q \, ,  \label{MQREN} \nonumber \\
& Z_M(\omega) = [ Z_P^2 Z_{S^0}^{-2} \cos^2\omega + \sin^2\omega ]^{1/2} 
\, . 
\end{eqnarray} 
$Z_{P}$ and $Z_{S^0}$ are the (mass--independent scheme)
renormalization constants of $\bar\chi \gamma_5 \tau_a \chi$
and $\bar\chi \chi$.

\subsection{Critical mass} 
\label{sec:MC}

For given values of $g_0^2$ and $r$, the appropriate value
of $M_{\rm cr}$ can be determined by adjusting $m_0$
so as to enforce one of the properties 
(chiral WTI's, pions with minimal mass) dictated by 
flavour chiral symmetry~\footnote{
In perturbation theory all the conditions of this type defining 
$M_{\rm cr}$ lead to a unique result: in fact O($ap$) terms can
be singled out and removed, since all momentum scales $p$ are
controllable at fixed $g_0^2$. Non-perturbatively this is no
longer the case, because there is no way of letting $a \Lambda_{QCD} 
\to 0$ at fixed $g_0^2$. At $g_0^2 > 0$
different definitions of $M_{\rm cr}$ in general differ 
by amounts of order $a \Lambda_{QCD}$.}. 
In the quark basis~(\ref{TWCHI}) and for any $m_2$-value, a sensible 
condition defining $M_{\rm cr}$ is given e.g.\ by (index $b=1,2$ only)
\beq
\sum_{\bf x} \partial_\mu \langle (\bar \chi \gamma_\mu \gamma_5 \tau_b \chi)(x)
{\cal O }(y,...) \rangle|_{m_0=M_{\rm cr}} = 0 
\, , \label{MCAWI}
\eeq
with ${\cal O }$ a conveniently chosen multilocal operator and
$ x \neq \{y,...\}$. Owing to the symmetry~(\ref{TILPAR_sp}),
numerical estimates of $M_{\rm cr}$ corresponding
to different values of $m_2$ differ by O($am_2^2$) from each other.
Since, by taking e.g.\
${\cal O } = (\bar \chi \gamma_5 \tau_b \chi)(y)$, 
the parameters $m_q$ and $\omega$
enter the condition~(\ref{MCAWI}) only through $m_2$,
one sees that, up to irrelevant O($a$) terms, $M_{\rm cr}$ is 
independent of $m_q$ and $\omega$. 
%
%

{\small \vspace{0.2cm}

In infinite volume, correlation functions and derived quantities may 
depend (e.g.\ if chiral symmetry is spontaneously broken) also on the path 
along which the critical point $(m_0,m_2) = (M_{\rm cr},0)$ is approached, 
and thus on $\omega$. However, since  renormalization is a local procedure, any
possible $\omega$-dependence of whatever determination of $M_{\rm cr}$ based 
on infinite volume quantities is necessarily limited to irrelevant 
O($a \Lambda_{QCD}$) contributions. 

Moreover, the combination 
of the spurionic invariances~(\ref{R5_sp2}) and~(\ref{TILPAR_sp}) 
implies that, if $m_0=M_{\rm cr}$ fulfills a certain condition 
defining the critical mass for a given $r$ (and $m_2$), then 
$m_0 = -M_{\rm cr}$ satisfies the same condition for $-r$ (and the same $m_2$).
The critical mass counterterm is thus an odd function of $r$:
\beq
M_{\rm cr}(-r) = -M_{\rm cr}(r) \, . \label{MCODD}
\eeq

If there exists an interval $[s_1,s_2]$ of $m_0$-values
for which the condition defining $M_{\rm cr}(r)$ is satisfied, 
the invariances~(\ref{R5_sp2}) and~(\ref{TILPAR_sp}) imply that the 
$m_0$-values in the interval $[-s_2,-s_1]$ are solutions of the same 
condition for $-r$.
In finite volume, analiticity in $m_0$ excludes the possible existence of
such intervals of solutions. They may however show up, though with
a width vanishing as $a \to 0$, if $M_{\rm cr}$ is defined through 
some infinite volume quantity, as for instance when it is determined by the  
vanishing of the charged pion mass and the ``Aoki phase scenario''~\cite{A_SS} 
is realized~\footnote{In this scenario the intervals of $m_0$-values  
with zero charged pion mass are expected to have O($a^2$) widths.}. 

\vspace{0.2cm} } 

Once a definition of $M_{\rm cr}(r)$ has been chosen for,
say, $r>0$, one must (and, as shown above, always
can) take $M_{\rm cr}(-r) = -M_{\rm cr}(r)$.
Otherwise one would unnecessarily spoil --by an artificial choice of
the untwisted mass counterterm-- the spurionic symmetry 
${\cal R}_5 \times \tilde{\cal P} \times (r \to -r) \times (m_0 \to - m_0)$
enjoyed by the lattice theory prior to renormalization. For these
reasons the criticism to eq.~(\ref{MCODD}) raised in ref.~\cite{AB}
is unjustified.

%
%
%
%
%

\section{Mass non-degenerate flavours}

For the case of $\omega = \pi/2$ (maximal twist),
the fermionic action of an $SU_f(2)$ pair of mass
non-degenerate quark is conveniently written~\cite{CAIRNS} as
\begin{eqnarray}
&S_{\rm{Fnd}}^{(\pi/2)}[\psi,\bar\psi,U]=
a^4 \sum_x\,\bar\psi(x)\Big{[}\gamma\cdot\widetilde\nabla+
\label{PHYSCHIND}\\
&-i\gamma_5\tau_1 W_{\rm{cr}}(r)
+m_q-\tau_3\epsilon_q\Big{]}\psi(x)
\, ,\nonumber
\end{eqnarray}
where to keep the mass term real and flavour diagonal I have used the matrix
$\tau_3$ to split the masses of the members of the doublet. Consequently the
Wilson term has been twisted with the flavour matrix $\tau_1$. 
Note that $m_q$ and $\epsilon_q$ are both real.

It has been shown~\cite{CAIRNS} that the quark mass splitting 
$\epsilon_q$ is only multiplicatively renormalized and 
the continuum flavour chiral WTI's can be implemented up to O($a$),
provided that~\footnote{$Z_S$ is the renormalization constant
of $\bar \chi \tau_a \chi$, $a=1,2,3$.}
\begin{eqnarray}
&&\hat{m}_q^{(-)}=\hat{m}_q-\hat{\epsilon}_q=Z_P^{-1}m_q-Z_S^{-1}\epsilon_q
\, ,\label{RMM} \\
&&\hat{m}_q^{(+)}=\hat{m}_q+\hat{\epsilon}_q=Z_P^{-1}m_q+Z_S^{-1}\epsilon_q
\label{RMP}
\end{eqnarray}
are identified as the renormalized (current) masses of the quarks in the doublet.

Remarkably, the fermionic 
determinant~\footnote{Here $D_{\rm Fnd}$ denotes the Dirac operator 
(a $2\times 2$ matrix in flavour space) corresponding
to the fermionic action~(\ref{PHYSCHIND}).}, 
\begin{eqnarray}
& {\rm det}[D_{\rm{Fnd}}^{(\pi/2)}] \geq
{\rm det}[Q_{\rm cr}^2+m_q^2-\epsilon_q^2] \, ,
\label{ND_DET} \\
& Q_{\rm cr} \equiv \gamma_5[\gamma \!\cdot\! \widetilde{\nabla}
+W_{\rm cr}(r)] = Q_{\rm cr}^\dagger \, .
\label{QCR}
\end{eqnarray}
is real and
positive, as long as $\epsilon_q^2 < m_q^2$~\cite{CAIRNS}.
This allows for unquenched Monte Carlo simulations, for instance by means
of algorithms of the multiboson or PHMC type based on
some polynomial approximation of $[D_{\rm Fnd}^\dagger D_{\rm Fnd}]^{-1/2}$.
 
Alternatively, lattice formulations of QCD with $N_f$ mass 
non-degenerate quarks of the (twisted) Wilson type can be 
obtained by taking the fermionic action of the form~\cite{OS,KSS,FR2}:
\begin{eqnarray}
 & S_F^{OS} \;\; = \;\; \sum_{f=1}^{N_f} \Big{\{} \; a^4 \sum_x \; \bar{q}_f(x) 
\nonumber \\
 & [ \gamma \!\cdot\! \widetilde{\nabla} + e^{-i\gamma_5\theta_f}
  W_{\rm cr}(r_f) + m_f ] q_f(x) \; \Big{\}} \, ,
\label{ND_DIA}
\end{eqnarray}
with $\theta = \sum_f \theta_f =0$~\footnote{The condition $\theta =0$ 
is necessary for the corresponding continuum limit theory be QCD with 
vanishing ``$\theta$-term''.} and $W_{\rm cr}(r_f)$
defined as in eq.~(\ref{WDONDM}). For all flavours $f$,
the critical mass 
$M_{\rm cr}(r_f)$ can be taken independent of $\theta_f$ 
and is given by the same dimensionless function $aM_{\rm cr}(r)$ 
as in the case of (un)twisted mass-degenerate
Wilson quarks~\cite{FR2,PSV}. The renormalized counterpart of the bare
quark mass $m_f$ reads~\cite{OS,FR2,PSV}
\beq  \label{MR_OS}
\hat{m}_f = Z_m(r_f) m_f \, , \quad\quad f=u,d,s,c,\dots \, .
\eeq 

Choosing $\theta_f \neq 0$ (at least for $u$ and $d$ quarks) 
allows to get rid of spurious quark zero modes, while keeping 
the action diagonal in flavour space. 
However, even if $\theta=0$, for generic values of the $m_f$'s
the lattice fermionic determinant corresponding to the action~(\ref{ND_DIA}) 
is complex~\cite{OS,KSS,PSV}. Actions of the type~(\ref{ND_DIA})
can thus be used in general only for valence quarks (see 
sect.~\ref{sec:WME}). Particular cases with positive quark determinant
exist: e.g.\ the $N_f=4$ case with 
$m_u = m_d$, arbitrary $m_s$ and $m_d$, $\theta_u = -\theta_d =
\pi/2$ and $\theta_s = \theta_c=0$~\cite{PSV}.

\section{Continuum and chiral limits}

In tmLQCD the relative magnitude of the lattice cutoff effects arising
from the chiral violating action terms may change significantly as a function
of the quark mass. This remark is particularly relevant in the regime
where chiral symmetry is spontaneously broken. In this situation,
in the continuum theory the chiral phase of the vacuum is driven 
by the phase of the quark mass term. The same must be true
on the lattice, thus ideally the continuum limit should be taken 
--at non-zero $\hat{m}_q$-- before letting $\hat{m}_q\rightarrow 0$. 

\subsection{O(a) improvement}

From an analysis \`a la Symanzik~\cite{SYM} of the leading cutoff effects
in tmLQCD (eq.~(\ref{PHYSCHI})) and by exploiting the invariances~(\ref{R5Dd}) 
and~\footnote{The spurionic symmetry ${\cal R}_5^{sp}$ holds for generic  
$\omega$ and follows directly from the symmetry~(\ref{R5_sp2}) using 
(\ref{BQMP}) and~(\ref{MCODD}).}  
\beq
{\cal R}_5^{sp} \equiv {\cal R}_5 \times (r \to -r) 
\times (m_q \to -m_q) \, ,
\label{R5_sp}
\eeq
it follows~\cite{FR1} that the expectation values of multilocal,
gauge invariant, multiplicatively renormalizable (m.r.) operators $O$
satisfy
\begin{eqnarray}
\hspace{-0.3cm}& \langle O(x_1,...,x_n) \rangle^{(\omega)}_{(r,m_q)}
+
\langle O(x_1,...,x_n) \rangle^{(\omega)}_{(-r,m_q)} =
\nonumber \\
\vspace{0.3cm}
\hspace{-0.3cm}& = 2 \zeta_{O}^{O}(r;\omega) \langle O(x_1,...,x_n)
\rangle_{(m_q)}^{\rm cont}
+ {\rm O}(a^2) \, . \quad \label{WA}
\end{eqnarray}
O($a$) effects cancel~\footnote{The same is true also for
O($a^{2k+1}$) effects ($k$ integer).} in the average of results 
obtained with opposite values of $r$ (Wilson average).

In case of maximal twist, $\omega = \pm \pi/2$, one can obtain
O($a$) improved results even from one single simulation at a given 
$r$-value. In fact invariance of the action 
under ${\cal P} \times (r \to -r)$~\footnote{
This symmetry follows directly from eqs.~(\ref{PAROP}) and~(\ref{MCODD}).}   
implies
\beq
\langle O(x_1,...) \rangle^{(\pm\pi/2)}_{(-r,m_q)} =
\eta_O \langle O({x_1^P,...}) \rangle^{(\pm\pi/2)}_{(r,m_q)}
\, , \label{SPEC}
\eeq
with $\eta_O$ the formal parity of $O$,
from which the second term in the l.h.s.\ of~(\ref{WA}) 
can be obtained.

As eq.~(\ref{WA}) holds for arbitrary space-time separations,
all the quantities derived from m.r.\ lattice correlators are free
from O($a$) cutoff effects, once the Wilson average, or
-- if $\omega = \pm \pi/2$ -- the average over opposite
values of all external three-momenta (see eq.~(\ref{SPEC})),
has been taken.

The extension to mass non-degenerate quarks is straightforward, since
the symmetries~(\ref{R5Dd}) and (with obvious modifications)~(\ref{R5_sp}) 
remain valid. 

At variance with
the familiar Symanzik's programme for O($a$) improvement~\cite{SYM}, 
this method does not require the addition (and the determination) of 
action and operator counterterms.

%
{\small \vspace{0.2cm}

The statement that a certain lattice quantity is O($a$) improved 
simply means that in the renormalized theory it approaches its continuum 
limit with a rate (asymptotically) quadratic in $a$. Moreover, as far 
as one is interested in continuum QCD with massive pions,
the renormalized quark mass $\hat{m}_q \propto m_q$ must have a 
non-zero limit as $a\to 0$~\footnote{
If $a \to 0$ at $\hat{m}_q=0$ and fixed physical volume, no
spontaneous breaking of chiral symmetry occurs and chiral
breaking cutoff effects are expected to be harmless~\cite{FR1}.}.
Under this condition O($a$) improvement holds
irrespectively of the value of $\hat{m}_q$. The contrary result 
of ref.~\cite{AB} comes from the fact that there $\hat{m}_q$ is 
allowed to be an O($a$), or O($a^2$), quantity. 
The question of the magnitude of residual cutoff effects,
even if parametrically ${\rm O}(a^2)$, when $\hat{m}_q$ is 
numerically small is however important in practice (e.g.\
for extrapolations of results to $a=0$).

\vspace{0.2cm} } 

Large cutoff effects 
can arise when working on a coarse lattice at small
$m_q$-values, just because both the quark mass term proportional 
to $m_q$ and the Wilson term (which has a different chiral orientation 
if $\omega \neq 0$) contribute to the breaking of chirality (and in 
particular to the chiral phase of the vacuum). 
In general, to avoid large lattice artifacts one should work
with parameters such that $a \Lambda_{\rm QCD}^2 \ll \, m_q$, but 
for O($a$) improved quantities it is conceivable that in several cases
the relative magnitude of the dominant (as $a \to 0$) cutoff effects is 
just O($a^2\Lambda_{\rm QCD}^3/m_q$). Scaling tests are thus important
to assess the magnitude of cutoff effects on various observables as
a function of $\hat{m}_q$. 

\subsection{ChPT analyses and phase diagram}

{\small \vspace{0.2cm}
Lattice chiral perturbation theory (ChPT) is an expansion
in powers of the quark mass (or external momenta) and the lattice
spacing and provides an explicit representation of 
(lattice estimates of) physical observables in the Goldstone
boson sector, as well as of the chiral phase of the lattice vacuum, in
terms of a few low energy constants to be determined~\cite{OB}.
Such a representation of observables, though usually limited to O($m_\pi^4$),
can be very useful in extracting physical information from simulation data.
When applied to tmLQCD~\cite{MUETAL}, lattice ChPT
involves both the quark mass $\hat{m}_q$ and the (rescaled) twist 
angle~\footnote{Dependence on $\hat{\omega}$ disappears as
$a \to 0$ at fixed $\hat{m}_q$.} 
$ \hat{\omega} = \tan^{-1}[ Z_{S^0}Z_P^{-1} \tan \omega] $, or equivalently
\beq \label{RECUMA}
\hat{m}_1 \equiv \hat{m}_q \cos \hat{\omega} \, , \quad\quad
\hat{m}_2 \equiv \hat{m}_q \sin \hat{\omega} \, .
\eeq 
Here I shortly summarize the main results obtained
to O($m_\pi^4$), including O($a$)~\cite{MUETAL} and
O($a^2$)~\cite{MUE,SHWU,SCO} artifacts,
for tmLQCD in different regimes. \\
} 

\vspace{-0.3cm}
{\bf Regime} ${\bf \hat{m}_q \gg {\rm O}(a)}$: The chiral phase of the
vacuum is determined by the term $\propto \hat{m}_q$, up to negligible 
higher order corrections.
O($a$) cutoff effects on pion masses and decay constants cancel in Wilson 
averages at generic $\hat\omega$ and are automatically absent at $
\hat\omega = \omega = \pm\pi/2$. The pion mass splitting is O($a^2$), as
expected. Lattice artifacts of order $a^2/\hat{m}_q$ however arise
from O($a$) terms in the chiral Lagrangian, e.g.\ in the pion
masses~\cite{SCO}. \\

\vspace{-0.3cm}
{\bf Regime} ${\bf \hat{m}_q \sim {\rm O}(a)}$: Once the O($a$) corrections to the
untwisted quark mass (related to the intrinsic non-perturbative uncertainty on
$M_{\rm cr}$) are taken into account, the chiral phase of the vacuum is
still determined by the term $\propto \hat{m}_q$ (up to small corrections). 
The concept of O($a$) improvement is of course no longer applicable. \\

\vspace{-0.3cm}
{\bf Regime} ${\bf \hat{m}_q \sim {\rm O}(a^2)}$: After taking into account
the O($a$) corrections to the untwisted quark mass, one finds that O($a^2$)
and O($\hat{m}_q$) terms in the chiral Lagrangian compete with each other
in determining the chiral phase of the vacuum. Depending on the sign of
the coefficient, $c_2$, of the term $\propto a^2$ in the chiral effective 
potential, one finds two possible scenarios~\footnote{Higher order
corrections are unlikely to alter the qualitative features
of these scenarios~\cite{MUE,SHWU}.}, which extend to twisted mass
$\hat{m}_2 \neq 0$ the ``Aoki phase'' ($c_2 >0$) and ``normal'' 
($c_2 <0$) scenarios of ref.~\cite{A_SS}. 
For $c_2>0$, the Aoki phase transitions are washed out 
into a crossover if $\hat{m}_2$ is non-zero. 
For $c_2<0$, the first order phase transition in the untwisted
mass parameter $\hat{m}_1$ that occurs at $(\hat{m}_1,\hat{m}_2)=(0,0)$ 
extends itself in the twisted mass plane, ending with two symmetrical 
second order points, $(\hat{m}_1, \hat{m}_2) = (0, \pm \hat{m}_{\rm 2c})$,
where the neutral pion mass vanishes. Note that $\hat{m}_{\rm 2c} \sim |c_2| a^2$.
The value of $c_2$, which depends on (un)quenching, 
$g_0^2$ and many details of the lattice action, is closely related 
to pion mass splitting: $m_{\pi^0}^2 - m_{\pi^\pm}^2 \propto c_2 a^2$, with 
positive proportionality constant.

\section{Numerical investigations}

A first convincing numerical evidence in favour of O($a$) improvement via
Wilson average and its remarkable simplicity in the case of tmLQCD with $\omega
= \pm \pi/2$ was obtained in ref.~\cite{JSUW}. There the scaling behaviour of the 
vector meson mass and the pseudoscalar decay constant was studied in quenched tmLQCD 
with two mass degenerate quarks for lattice spacings in the range $0.068 \div 0.123$~fm.
All quantities were expressed in units of Sommer's scale $r_0$
and, for each value of $g_0^2$, $m_0$ was set to $M_{\rm cr}$ and
$m_2$ was chosen so as to obtain $r_0m_{\rm PS}=1.79$. 
The results were also compared with the corresponding 
ones for plain and clover-improved Wilson quarks.

{\small \vspace{0.2cm}
An extension of this scaling study to lower quark masses (down to 
$r_0m_{\rm PS} \sim 0.6$) and few more 
lattice spacings is currently in  progress~\cite{SHETAL} and should
provide useful insights on the size of residual cutoff effects as a function
of the quark mass. A comparative study of several hadronic observables 
for a wide range of quark masses, using tm and overlap fermions in the  
quenched approximation at $\beta=5.85$, has also been 
presented~\cite{PAETAL}. Given the different chiral properties of tm
and overlap fermions, information on the continuum limit, coming 
from scaling tests, such as those of ref.~\cite{JSUW,SHETAL}, 
may be very beneficial for the interpretation of the results. 
The performance of linear solvers for quark propagators in the 
two different lattice formulations has also been studied at 
$\beta=5.85$ for given volumes and pion masses~\cite{JAETAL}. 

A nice computation of the pion form factor, $F(Q^2)$,
has been carried out in quenched tmLQCD at $\omega = \pi/2$,
$\beta=6.0$ and pion masses of about 660~MeV and 470~MeV~\cite{ARL}.
The conserved (one-point split) isotriplet vector current 
$V_\mu^3$~\cite{TM} was employed and the matrix element 
\beq
\langle \pi^+({\bf p}_{f}) | V_0^3 | \pi^+({\bf p}_{i}) \rangle
= F(Q^2) [E({\bf p}_{f})+E({\bf p}_{i})] \, ,
\eeq
with $Q = p_f - p_i$,
was O($a$) improved by averaging over opposite 
values of ${\bf p}_{i,f}$~\cite{FR1}. Results agree with meson vector
dominance and other O($a$) improved computations and cover
a wide range of $Q^2$-values.
} 

\vspace{0.2cm}
The phase diagram of unquenched tmLQCD with two mass degenerate quarks
(action~(\ref{TWISCHI})) in the plane ($m_0 , m_2$)
is currently under study as a function of $g_0^2$ and
for different choices of the pure gauge action (plaquette and DBW2).
Results with the plaquette YM action at $\beta=5.2$ did reveal the 
presence of metastabilities in several observables (plaquette, pion
mass, $m_\chi^{\rm PCAC}$)~\footnote{See eq.~(12) of ref.~[5a]
for the definition of $m_\chi^{\rm PCAC}$.}
for small values of $m_2$ and $m_0$ close
to its critical value (the value where $m_\chi^{\rm PCAC}$ 
should vanish)~\cite{FARCH}. 

{\small \vspace{0.1cm}
Long living metastable states (associated with 
values of $m_\chi^{\rm PCAC}$ of different sign) were 
identified and the metastability in the plaquette was related
to that in the untwisted condensate $\langle \bar\chi \chi \rangle$.
These findings were interpreted as evidence for a
segment of first order phase transition, i.e.\ 
the scenario arising from lattice 
ChPT in the regime $\hat{m}_q \sim {\rm O}(a^2)$ for $c_2<0$.
Metastabilities in unquenched simulations with Wilson-like quarks would
thus be ultimately related to chiral violating cutoff effects and should
be absent for $m_2 \neq 0$, if $c_2>0$, and for $|m_2| > m_{\rm 2c} \sim |c_2|a^2$, 
if $c_2<0$. Whenever $c_2 <0$, the condition $|m_2| > m_{\rm 2c}$ 
can always be fulfilled by sufficiently decreasing $a$,
but it would of course be desirable to
have both $c_2$ small in modulus and a good $a^2$-scaling behaviour already 
for $a \sim 0.10 \div 0.15$~fm. To what extent this can be achieved by simple 
modifications of the irrelevant terms of the lattice action is still an 
open question.
} 

\section{Weak matrix elements and tm quarks}
\label{sec:WME}

Although tm Wilson quarks do not preserve full chiral symmetry, 
they can be used to compute many weak matrix elements with no
or substantially reduced (with respect to standard Wilson fermions)
operator mixings. To achieve this remarkable result,
the key step is to choose the details of the UV regularization
of the (typically four quark) operators on a case by case basis.
More precisely, one can conveniently employ for
the valence quarks in the operator of interest a flavour
diagonal action of the form~(\ref{ND_DIA}) and choose the value
of the angles $\theta_f$
so as to maximally simplify operator renormalization. 
If the UV regularization chosen for the quarks in the
operator (and the hadron interpolating fields) does not admit
a positive defined fermionic determinant, in unquenched studies
one can adopt a different regularization for the sea quarks, e.g.\
that provided by maximally twisted Wilson quarks (see eq.~(\ref{PHYSCHIND})).
In this case, to make contact, as $a \to 0$, with continuum 
QCD~\footnote{Here I am interested in the theory with massive quarks.},
the masses of the sea and valence quarks of the same flavour must be
matched, for instance by imposing the equality of the renormalized
current quark masses~\cite{FR2}. 

Within the quenched approximation, use of tm Wilson quarks 
and non-perturbative renormalization (in the SF scheme) has recently
led to a precise computation of $B_K$, with no mixings and
extrapolation of results to the continuum limit~\cite{DHPSV}. 

Moreover it has been remarked~\cite{PSV} that by using the clover
improved version of the valence
quark action~(\ref{ND_DIA}), with $N_f=4$, $\theta_u = -\theta_d =
\pi/2$ and either $\theta_s = - \theta_c = \pi/2$ or
$\theta_s = \theta_c = 0$, 
the renormalization of $K \to \pi$ matrix elements requires
at most linearly divergent counterterms, which can be determined by
enforcing parity.

A rather general strategy to make use of maximally twisted Wilson 
quarks for evaluating (un)quenched weak matrix elements with 
neither wrong chirality operator mixings nor O($a$) cutoff effects
has been presented in ref.~\cite{FR2}. Besides using different (cleverly
chosen) tm Wilson regularizations for sea and valence quarks, the approach is
based on the remark that renormalizable lattice models with four sea 
quarks and a certain number of (possibly replicated) valence quarks, 
plus corresponding ghosts, yield, among others, operator matrix elements 
that coincide --in the continuum limit and provided the renormalized sea and valence
quark masses are appropriately matched-- with those of $N_f=4$ Euclidean QCD.
The method was illustrated in ref.~\cite{FR2} by discussing the evaluation of
$B_K$ as well as $K \to \pi\pi$ and $K\to\pi$ amplitudes. In ref.~\cite{MDM}
it has been pointed out that it can be 	
extended to static quarks and applied
e.g.\ to the computation of $B_B$.

\section{Conclusions}

Lattice formulations of QCD with chirally twisted Wilson
quarks provide a framework for non-perturbative numerical studies
where spurious quark zero modes are absent and O($a$) cutoff effects,
whenever present, can be removed either \`a la Symanzik or
by a new simpler method. The magnitude of the residual O($a^2$)
scaling violations in the small quark mass region is under investigation.
In unquenched simulations a non-trivial phase diagram is also expected,
and numerical studies of it are in progress. 
Finally, tmLQCD can be in various ways extended to 
mass non-degenerate quarks and adapted so as to make possible
computations of matrix elements
of the weak effective Hamiltonian with reduced or no mixings.

\section*{Acknowledgements}
I thank the LOC of Lattice~2004 for the stimulating
atmosphere of the conference, 
G.C.~Rossi for critical reading of the manuscript,   
G.~M\"unster, S.~Sharpe and P.~Weisz for valuable discussions.

\end{document}